\documentclass[12pt]{article}
\usepackage{epsfig,amsfonts,amssymb}
\begin{document}

\newcommand{\be}{\begin{equation}}
\newcommand{\ee}{\end{equation}}
\newcommand{\bea}{\begin{eqnarray}}
\newcommand{\eea}{\end{eqnarray}}
\newcommand{\beas}{\begin{eqnarray*}}
\newcommand{\eeas}{\end{eqnarray*}}

\baselineskip 14 pt

\begin{titlepage}
\begin{flushright}
{\small CU-TP-1072} \\
{\small hep-th/0211124}
\end{flushright}

\begin{center}

\vspace{5mm}

{\Large \bf Brane gas cosmology in M-theory: \\ late time behavior}

\vspace{3mm}

Richard Easther${}^a$\footnote{\tt easther@physics.columbia.edu},
Brian R.~Greene${}^{ab}$\footnote{\tt greene@physics.columbia.edu},
Mark G.~Jackson${}^c$\footnote{\tt markj@physics.columbia.edu} \\
and Daniel Kabat${}^c$\footnote{\tt kabat@physics.columbia.edu}

\vspace{2mm}

${}^a${\small \sl Institute for Strings, Cosmology and Astroparticle
Physics} \\
{\small \sl Columbia University, New York NY 10027}

\vspace{1mm}

${}^b${\small \sl Department of Mathematics} \\
{\small \sl Columbia University, New York, NY 10027}

\vspace{1mm}

${}^c${\small \sl Department of Physics} \\
{\small \sl Columbia University, New York, NY 10027}

\end{center}

\vskip 0.3 cm

\noindent
We investigate the late-time behavior of a universe containing a
supergravity gas and wrapped 2-branes in the context of M-theory
compactified on $T^{10}$.  The supergravity gas tends to drive uniform
expansion, while the branes impede the expansion of the directions
about which they are wrapped.  Assuming spatial homogeneity, we study
the dynamics both numerically and analytically.  At late times the
radii obey power laws which are determined by the brane wrapping
numbers, leading to interesting hierarchies of scale between the
wrapped and unwrapped dimensions.  The biggest hierarchy that could
evolve from an initial thermal fluctuation produces three large
unwrapped dimensions.  We also study configurations corresponding to
string winding, in which the M2-branes are all wrapped around the
(small) 11th dimension, and show that this recovers the scenario
discussed by Brandenberger and Vafa.

\end{titlepage}

\section{Introduction}
\label{Intro}

The dual realizations that D-branes are an intrinsic component of
string theory \cite{dbranes} and that the five distinct vacua of
string theory are actually different limits of eleven-dimensional
M-theory \cite{witten} have had dramatic implications for cosmology.
Both developments widen the range of options open to model-builders,
but they also increase the number of constraints that a
successful ``string'' cosmological model must satisfy.  In particular,
branes have given rise to a huge number of new models, which typically
postulate that our apparently three dimensional universe is actually a
3-brane embedded in a higher dimensional background \cite{braneworld}.
Even in ``standard'' cosmological models, however, branes are expected
to be in thermal equilibrium with other matter in the early universe,
and must therefore be incorporated into any complete cosmological
scenario.

{\em Brane gas cosmology\/} is devoted to understanding the
cosmological role of the ``gas'' of branes that inhabited the early
universe. A particular focus of this program has been whether brane
gas cosmologies single out a special number of ``large'' dimensions.
In 1989, Brandenberger and Vafa argued that cosmological models
containing a gas of winding strings can produce, at most, three
macroscopic dimensions \cite{bv}.  This work has been tested and
extended in a number of ways since then \cite{TseytlinVafa,Tseytlin,
Sakellariadou,AlexanderBrandenbergerEasson,BrandenbergerEassonKimberly,
Easson,EastherGreeneJackson,WatsonBrandenberger}. The key observation
underlying this work is that extended, $p$ dimensional objects have
$p+1$ dimensional worldvolumes, and from a purely topological
perspective these worldvolumes will only intersect in $2(p+1)$ 
spacetime dimensions or less.  Thus, strings can only interact in three
spatial dimensions (or less), whereas 2-branes can find each other in
up to five spatial dimensions. One obviously hopes that Brandenberger
and Vafa's conclusion that strings naturally lead to a three
dimensional universe is not undermined when branes are added to the
picture. Indeed Alexander, Brandenberger and Easson
\cite{AlexanderBrandenbergerEasson} have argued that when branes are
included, strings will still dominate the evolution of the universe at
late times, so the basic Brandenberger-Vafa mechanism survives.

This paper extends brane gas cosmology in two important ways. First,
rather than work with ten dimensional string theory, we take
eleven-dimensional M-theory as our starting point.  This is a natural
basis for any theory that hopes to understand the origin of the three
large dimensions, since string theory is itself a compactification of
M-theory, with the compact direction providing the string dilaton
\cite{witten}.  Consequently, string theory implicitly assumes that
one direction is already on a different footing from the rest, and in
the long run we would hope to explain this rather than inject it as a
hypothesis. Second, rather than using solely thermodynamic arguments,
we study the dynamical evolution of an M-theoretic universe containing
M2-branes and a gas of supergravity particles. We ignore the M5-branes
that are present in the M-theory spectrum, since they can annihilate
in the full 11 dimensional M-theory spacetime.  We are particularly
interested in cases where the different directions (assumed to be
toroidally compactified) are anisotropically wrapped by M2-branes. We
develop the equations of motion for the general case in which each of
the 10 spatial dimensions is distinguishable from the others, and
describe the configuration of wrapped branes and anti-branes with a
``wrapping matrix''.  Although we assume a toroidal topology,
numerical and analytical results \cite{EastherGreeneJackson} indicate
that wrapping dynamics should be insensitive to topology, and hence we
believe our conclusions extend to more realistic compactifications.

Strings are obtained when one dimension of the M2-branes gets wrapped
on the 11th dimension, which is taken to be much smaller than the
Planck scale.  Similarly, D-branes in string theories can be obtained
as various configurations of M2- and M5-branes.  Thus we no longer
need to consider five separate brane gas cosmology scenarios, one for
each string theory: our goal is to see if we can use M-theoretic
models to make generic predictions about the overall form of the
universe.

Motivated by the original Brandenberger-Vafa scenario, we focus on the
late time asymptotics of universes with a subspace that has no
wrapping modes, presenting both an analytic discussion and specific
numerical solutions.  Here we are primarily concerned with what is
{\em possible\/}, rather than what is {\em probable\/}, and focus on
the most interesting initial states that are permitted by general
topological arguments. In a subsequent paper we will discuss the
thermodynamics of a brane dominated universe in more detail, in order
to establish the relative likelihood of different initial
states. Consequently, we pay most attention to the following scenario.
We imagine that thermal fluctuations drive subspaces of various
dimensions to momentarily expand to a larger than average size. If the
fluctuation involves a five dimensional subspace, interesting
nontrivial dynamics can ensue. Fully wrapped 2-brane/anti-2-brane
pairs will generically annihilate in five space dimensions, thereby
removing their restoring force. This leaves partially wrapped 2-branes
that have one dimension wrapped along the expanding five-space and one
dimension wrapped along one of the other, smaller dimensions. From an
effective five-dimensional perspective, such branes appear to be
wrapped one-dimensional objects -- strings -- and naively appear to
impede further expansion. Yet, we imagine that subsequent thermal
fluctuations within this five-dimensional environment will, once
again, drive various subspaces to expand further, and if such a
subspace should happen to have three (or fewer) dimensions, the
effective wrapped strings will be able to annihilate, thus allowing
the subspace to expand without further constraint.  We therefore see
the possibility of generating an interesting hierarchy with a large
three dimensional subspace emerging from an intermediate five
dimensional subspace, with all remaining dimensions small
\cite{AlexanderBrandenbergerEasson}. One of our goals is to study this
scenario in detail to determine if the dynamics leverages the
topological reasoning we have used and drives a hierarchy between the
dimensions.

In Section 2 we derive the equations of motion and discuss some of
their general properties. In section 3 we present numerical solutions
to the equations of motion, and in section 4 we study the behavior at
late times analytically.  In section 5 we consider string winding, to
make contact with the original scenario of Brandenberger and Vafa.  In
section 6 we discuss the implications of our results, and outline
possibilities for future work.

\section{Homogeneous brane gas dynamics}

Our models are governed by M-theory which is well-described by
eleven-dimensional supergravity when the radii and curvature scales
are larger than the eleven-dimensional Planck length.  This assumption will be
justified in hindsight, as the cosmologically relevant solutions we
obtain contain growing radii.  We consider two types of matter.  The
first is the massless supergravity spectrum consisting of 128 bosonic
and 128 fermionic degrees of freedom -- we ignore massive modes since
we expect that these will quickly decay.  The second ingredient is
wrapped M2-branes.  Although M-theory also contains M5-branes, these
will annihilate quickly in 10+1 dimensions and thus should not be
significant to the late-time behavior.  We assume that the M2-branes
are at rest and ignore fluctuations on the brane volume, as these
effects are subleading in our large radii assumption.  Finally, to
make the analysis tractable, we assume that the particle and brane
gases are homogeneous.

Using the metric ansatz
\be
ds^2 = - dt^2 + \sum_{i=1}^d \bigl(R_i(t)\bigr)^2 d\theta_i^2
\qquad 0 \leq \theta_i \leq 2\pi
\ee
the non-vanishing Christoffel symbols are
\be
\Gamma^t_{ij} = \delta_{ij} R_i \dot{R}_i, \qquad \Gamma^i_{tj} =
\Gamma^i_{jt}   = \delta^i{}_j
{\dot{R}_i \over R_i},
\ee
while the non-vanishing components of the Einstein tensor are
\bea
G^t{}_t & = & {1 \over 2} \sum_{k \not= l} {\dot{R}_k \dot{R}_l \over
R_k R_l},  \\
G^i{}_i & = & \sum_{k \not= i} {\ddot{R}_k \over R_k} + {1 \over 2}
\sum_{k \not= l} 
{\dot{R}_k \dot{R}_l \over R_k R_l} - \sum_{k \not= i} {\dot{R}_k
\dot{R}_i \over R_k R_i},
\eea
where there is no implied summation on $i$ in the second line.

We first introduce a gas of massless supergravity particles, with
energy density $\rho_{\rm S}$ and pressure $p_{\rm S}$.  For
simplicity we take the gas to be homogeneous and isotropic, with a
perfect fluid stress tensor
\be
T^\mu{}_\nu = {\rm diag}(-\rho_{\rm S},p_{\rm S},\ldots,p_{\rm S})\,.
\ee
The equation of state appropriate for $d$ spatial dimensions fixes
$p_{\rm S} = {1 \over d} \rho_{\rm S}$, while covariant conservation
of the stress tensor requires 
\be
\rho_{\rm S} = {{\rm const.} \over V^{(d+1)/d}}\,,
\ee
where the volume of the $d$ dimensional torus is simply
\be
V = \prod_i{2\pi R_i}\,.
\ee

The second source of stress-energy in our model universe is a gas of
2-branes, wrapped on the various cycles of the torus.  These are
characterized by a matrix of wrapping numbers $N_{ij}$, where we take
$N_{i<j}$ to represent the number of branes wrapped on the $(ij)$
cycle, while $N_{i>j}$ represents the number of antibranes.  A single
M2-brane is described by the Nambu-Goto action\footnote{See also a
similar discussion of the action of a p-brane in \cite{brandboehm}.}
\bea
S & = & - T_2 \int d^3 \xi \sqrt{- \det g_{\alpha\beta}} \\
g_{\alpha\beta} & = & \partial_\alpha X^\mu \partial_\beta X^\nu
g_{\mu\nu}
\eea
where the brane tension $T_2 = {1 \over (2\pi)^2 \ell_{11}^3}$.  This
leads to the stress tensor
\be
T^{\mu\nu} = - T_2 \int d^3\xi \delta^{11}(X - X(\xi)) \sqrt{- \det g} \,
g^{\alpha\beta} \partial_\alpha X^\mu \partial_\beta X^\nu\,.
\ee
For simplicity we assume that the branes are at rest, and ignore any
possible excitations on the brane worldvolumes.  Then, for a single
brane wrapped on the $(12)$ cycle and uniformly smeared over the
transverse $T^{d-2}$, the stress tensor is
\be
T^\mu{}_\nu = - {T_2 \over 2 \pi R_3 \cdots 2 \pi R_d} {\rm
diag}(1,1,1,0,\ldots,0)\,.
\ee
With a matrix of wrapping numbers, the non-zero components of the
brane gas stress tensor are
\bea
T^t{}_t & = & - {T_2 \over V} (2\pi)^2 \sum_{k \not= l} R_k  R_l N_{kl} \\
T^i{}_i & = & - {T_2 \over V} (2\pi)^2 \sum_{k \not= i} R_k  R_i
\left(N_{ki} + N_{ik}\right)
\eea
with no sum on $i$ in the second line.

We now insert these two sources of energy-momentum into the right hand 
side of the $d$-dimensional Einstein equations,
\be
G^\mu{}_\nu = - 8 \pi G T^\mu{}_\nu
\ee
where the gravitational coupling is
\be
16 \pi G = (2 \pi)^8 \ell_{11}^9 \,.
\ee
The time-time equation can be solved for the energy density and hence
pressure of the supergravity gas,
\be  \label{Constraint}
8 \pi G \rho_{\rm S} = {1 \over 2} \sum_{k \not=l} {\dot{R}_k
\dot{R}_l \over R_k R_l}
- {8 \pi G T_2 \over V} (2\pi)^2 \sum_{k \not= l}  R_k  R_l
N_{kl} \, .
\ee
After a little re-arrangement, the space-space equations yield the
following set of second-order differential equations for the radii.
\bea
{\ddot{R}_i \over R_i} & = & {8 \pi G T_2 \over V}
\left[{2d+1 \over d(d-1)} \,
(2\pi)^2 \sum_{k \not= l}  R_k  R_l N_{kl} -
(2\pi)^2 \sum_{k \not= i}  R_k  R_i \left(N_{ki} + N_{ik}\right) \right]
\nonumber \\
& & + {1 \over 2 d} \sum_{k \not= l} {\dot{R}_k \dot{R}_l \over R_k R_l}
- \sum_{k \not= i} {\dot{R}_k \dot{R}_i \over R_k R_i} \, .
\label{eom}
\eea

\subsection{Some exact solutions}

With no wrapped branes ($N_{ij} = 0$) the model reduces to an 11
dimensional radiation dominated universe, and it is easy to find a
number of exact solutions. In particular, with the ansatz
\be
\label{poweransatz}
R_i(t) = c_i \left(t - t_0\right)^{\alpha_i}
\ee
we find the following solutions to (\ref{eom}):
\begin{itemize}
\item
Flat spacetime, with $\alpha_1 = \cdots = \alpha_d = 0$, in which case 
$\rho_{\rm S}$ will vanish.
\item
Kasner solutions, with $\sum_i \alpha_i = \sum_i (\alpha_i)^2 = 1$.
The energy density $\rho_{\rm S}$ vanishes, so these are vacuum
solutions.  Note that the exponents $\alpha_i$ must lie in the
interval $[-1,1]$, and that at least one of the $\alpha_i$'s must be
negative.
\item
Radiation dominated, with $\alpha_1 = \cdots = \alpha_d = {2 \over d +
1}$.  The energy density of the radiation is non-zero, fixed by
the constraint (\ref{Constraint}).
\end{itemize}
One can also find a solution with brane wrapping but no gravitons, in
which we set $N_{ij} = n$ for all $i \not= j$ but take $\rho_{\rm S} =
0$.  Using the same ansatz (\ref{poweransatz}), but setting all the
$R_i$ equal (so that $c_i = c$ and $\alpha_i = \alpha$), demanding
that $\rho_{\rm S} = 0$ implies
\be
{\alpha^2 \over 2 (t - t_0)^2} = {8 \pi G T_2 n \over (2 \pi R)^{d-2}}\,.
\ee
This equation is solved for all times by setting $\alpha = {2 \over
d-2}$, and choosing the value of $c$ appropriately.  Given these
values, one can check that the $\ddot{R}_i$ equations (\ref{eom}) are
also satisfied.

The ansatz (\ref{poweransatz}) doesn't give the most general solution
to the equations of motion, since we ought to be able to specify $2d$
independent initial radii and velocities.

\subsection{General features}

To study the solutions of the field equations (\ref{eom}), it is
convenient to work in terms of new variables $\lambda_i(t) \equiv
\log{(2 \pi R_i(t))}$ introduced in \cite{TseytlinVafa}.  The field
equations become
\begin{equation}
\label{eom2}
\ddot{\lambda_i} + {\dot{\rm V} \over {\rm V}} \dot{\lambda}_i = 8 \pi
G \left({1 \over d}
\rho_{\rm S} + {3 \over d - 1} \rho_{\rm B}\right) - {8 \pi G T_2
\over V}
e^{\lambda_i} \sum_{j \not= i} (N_{ij} + N_{ji}) e^{\lambda_j}\,.
\ee
The volume of the spatial torus is
\be
V = e^{\sum_i \lambda_i}\,,
\ee
the energy density from brane tension is
\be
\rho_{\rm B} = {T_2 \over V} \sum_{i \not= j} e^{\lambda_i}
e^{\lambda_j} N_{ij}
\ee
and the energy density from supergravity excitations is fixed by the
constraint (\ref{Constraint})
\begin{equation}
\label{GttConstraint}
8 \pi G \rho_{\rm S} = {1 \over 2} \sum_{i \not= j} \dot{\lambda_i} \dot{\lambda_j}
- 8 \pi G \rho_{\rm B}.
\end{equation}

As noted in \cite{TseytlinVafa}, this system of equations can be
regarded as describing a non-relativistic particle moving in $d$
dimensions.  The particle has a coefficient of friction, given by
$\dot{V} / V$, due to the expansion of the universe.  A
position-dependent force acts on the particle, given by the right hand
side of (\ref{eom2}).  This force consists of two terms.  The first
term,
\be
\label{F1}
F_i^{(1)} = 8 \pi G \left({1 \over d} \rho_{\rm S} + {3 \over d - 1} \rho_{\rm B}\right)\,,
\ee
is positive definite and is the same for every value of $i$.  It
drives a uniform expansion of the universe.  The second term,
\be
\label{F2}
F_i^{(2)} = - {8 \pi G T_2 \over V} e^{\lambda_i}
\sum_{j \not= i} (N_{ij} + N_{ji}) e^{\lambda_j}\,,
\ee
is either zero or negative and it suppresses the growth of dimensions
with large wrapping numbers.  In the M-theory context, this is the
mechanism by which anisotropic wrapping numbers lead to an anisotropic
expansion of the universe.  Another way of stating this is to note
that the equations of motion (\ref{eom2}) imply that
\be
\ddot{\lambda}_i - \ddot{\lambda}_j + {\dot{V} \over V}(\dot{\lambda}_i - \dot{\lambda}_j) = 8 \pi G (p_i - p_j)
\ee
where $p_i$ is the pressure exerted on the $i{}^{\hbox{\em th}}$
dimension.  Thus differential pressures, of the sort exerted by an
anisotropic brane gas, lead to differential expansion rates.

Let us comment on some general properties of these
equations.\footnote{These results overlap with the work of Don Marolf
\cite{Don}.}  First note that the volume of the torus increases
monotonically with time.  This follows from rewriting the constraint
(\ref{GttConstraint}) in the form
\be
\left({\dot{V} \over V}\right)^2 = \left(\sum_i \dot{\lambda}_i\right)^2
= \sum_i (\dot{\lambda}_i)^2 + 16 \pi G
\left(\rho_{\rm S} + \rho_{\rm B}\right)\,.
\ee
The right hand side is positive definite for all nontrivial models,
vanishing only for the trivial case of periodically identified
Minkowski space.  We can choose the direction of time to make $\dot{V}
> 0$.

The rest of this paper is primarily concerned with models where the
wrapping matrix $N_{ij}$ is anisotropic.  Motivated by our discussion
in section 1, we assume that the spatial dimensions fall into three
classes.  We refer to a direction $i$ such that $N_{ij} = N_{ji} = 0$
for all $j$ as {\em unwrapped\/}. Directions $i$ for which $N_{ij}$
and $N_{ji}$ are nonzero except for those $j$ corresponding to an
unwrapped direction are referred to as {\em fully
wrapped\/}. Directions $i$ where some of the $N_{ij}$ or $N_{ji}$ are
zero for values of $j$ which are not unwrapped are referred to as {\em
partially wrapped\/}.  That is, listing dimensions in the order
unwrapped, partially wrapped, fully wrapped, the wrapping matricies we
will consider look like
\be
\label{WrappingMatrix}
\left(
\begin{array}{cccccccccc}
\cdot & 0 & 0 & 0 & 0 & 0 &0 & 0 & 0 & 0\\
0 & \cdot &  & 0 & 0 & 0 &0 & 0 & 0 & 0\\
0 & 0 & \cdot & 0 & 0 & 0 &0 & 0 & 0 & 0\\
0 & 0 & 0 & \cdot & 0 & N_{4,6} &N_{4,7} & N_{4,8} & N_{4,8} & N_{4,10}\\
0 & 0 & 0 & 0 & \cdot & N_{5,6} &N_{5,7} & N_{5,8} & N_{5,9} & N_{5,10}\\
0 & 0 & 0 & N_{6,4} & N_{6,5} & \cdot &N_{6,7} & N_{6,8} & N_{6,9} &
N_{6,10}\\
0 & 0 & 0 & N_{7,4} & N_{7,5} & N_{7,6} & \cdot & N_{7,8} & N_{7,9} &
N_{7,10}\\
0 & 0 & 0 & N_{8,4} & N_{8,5} & N_{8,6} &N_{8,7} & \cdot & N_{8,9} &
N_{8,10}\\
0 & 0 & 0 & N_{9,4} & N_{9,5} & N_{9,6} &N_{9,7} & N_{9,8} &\cdot &
N_{9,10}\\
0 & 0 & 0 & N_{10,4} & N_{10,5} & N_{10,6} &N_{10,7} & N_{10,8} & N_{10,9} &
\cdot
\end{array}
\right) \label{sample}
\ee 
The diagonal entries in the matrix are irrelevant.  In this example
the directions 1 through 3 are unwrapped, 4 and 5 partially wrapped,
and 6 through 10 fully wrapped (assuming all the $N_{ij}$ written out
explicitly in the above matrix are non-zero).

Now consider what happens when $m$ dimensions are unwrapped and $d-m$
dimensions are either partially or fully wrapped.  We introduce
\bea
\mu & = & \sum_{i = 1}^m \lambda_i = \log \,\, (\hbox{\rm
volume of unwrapped torus}), \\
\Lambda & = & \sum_{i = m+1}^d \lambda_i = \log \,\, (\hbox{\rm
volume of wrapped torus})\,.
\eea
Summing over the appropriate values of $i$, and using the definition
of $\rho_{\rm B}$, we find the following differential equations for
$\mu$ and $\Lambda$:
\begin{eqnarray}
\label{mu}
\ddot{\mu} + (\dot{\mu} + \dot{\Lambda}) \dot{\mu}
& = & 8 \pi G \left({m \over d} \rho_{\rm S} + {3 m \over d-1}
\rho_{\rm B}\right) \\
\label{Lambda}
\ddot{\Lambda} + (\dot{\mu} + \dot{\Lambda}) \dot{\Lambda}
& = & 8 \pi G \left({d-m \over d} \rho_{\rm S} + {d - 3 m + 2 \over
d-1} \rho_{\rm B}\right)
\end{eqnarray}
The overall character of the dynamics depends on $m$. For small $m$,
both terms in the right hand side for $\ddot{\Lambda}$ are positive,
and if $\dot{\Lambda}$ is zero, then the second derivative must be
positive, leading to a local minimum.  Conversely, with a larger value
of $m$, the right hand side of this equation has both a positive and a
negative term, and both local maxima and minima are allowed.  For
$\mu$ there can never be a negative term on the right hand side, so if
these directions are initially expanding they will then expand
forever.

When $d - 3m + 2$ vanishes the brane tension does not contribute to
the growth of the internal dimensions.  This picks out a special
dimensionality of the unwrapped subspace, for which the internal
dimensions have nearly constant radii.  When $d = 10$ this occurs for
$m = 4$.

Perhaps the most important feature of these equations, which we
analyze in more detail in section \ref{LateTime}, is that an attractor
mechanism governs the behavior at late times.  That is, the radii at
late times are determined solely by the wrapping matrix, and are (up
to some scaling symmetries discussed in section \ref{LateTime})
independent of the initial radii and velocities.  This is intuitively
clear from examining the second term in the ``force'', (\ref{F2}).
Suppose one of the radii, say $R_1$, starts out with an unusually
small (or large) value.  This does not affect the force on $R_1$
itself, since $R_1$ appears in both the numerator and denominator of
$F_1^{(2)}$.  But the force on the other radii will become larger (or
smaller), in such a way as to restore a balance between the different
radii.  This suggests that the behavior at late times is solely
determined by the wrapping matrix.

\section{Numerical results}

We now present numerical solutions to the equations of motion
(\ref{eom}).  The technique is to start with a set of initial radii
and velocities, then evolve the system both forwards and backwards in
time using a Runge-Kutta routine.  Motivated by our discussion in
section \ref{Intro}, we assume a wrapping matrix with $m_1$ unwrapped
dimensions, $m_2$ partially wrapped directions, and $m_3 = d- m_1-
m_2$ fully wrapped dimensions. For the specific case $m_1=3$ and $m_2
= 2$ in a ten dimensional universe we have the wrapping matrix given
by (\ref{sample}). The wrapping numbers should be chosen based on an
understanding of thermal fluctuations in the early universe.  Pending
this, for our numerical calculations we simply take the non-zero
wrapping numbers to be randomly chosen integers $N_{ij} \in \lbrace
1,2,3 \rbrace$.  We take the initial radii $R_i$ to be randomly chosen
between 1 and 5 Planck lengths, and we take the initial velocities
$\dot{R}_i$ to be randomly chosen between -0.5 and +0.5.  A typical
numerical solution for 3 unwrapped dimensions, 2 partially wrapped
dimensions and 5 fully wrapped dimensions, is shown in Fig.~1.  A
solution for $m_1 = 3$, $m_2 = 4$ and $m_3 = 3$ is shown in Fig.~2.

\begin{figure}
\epsfig{file=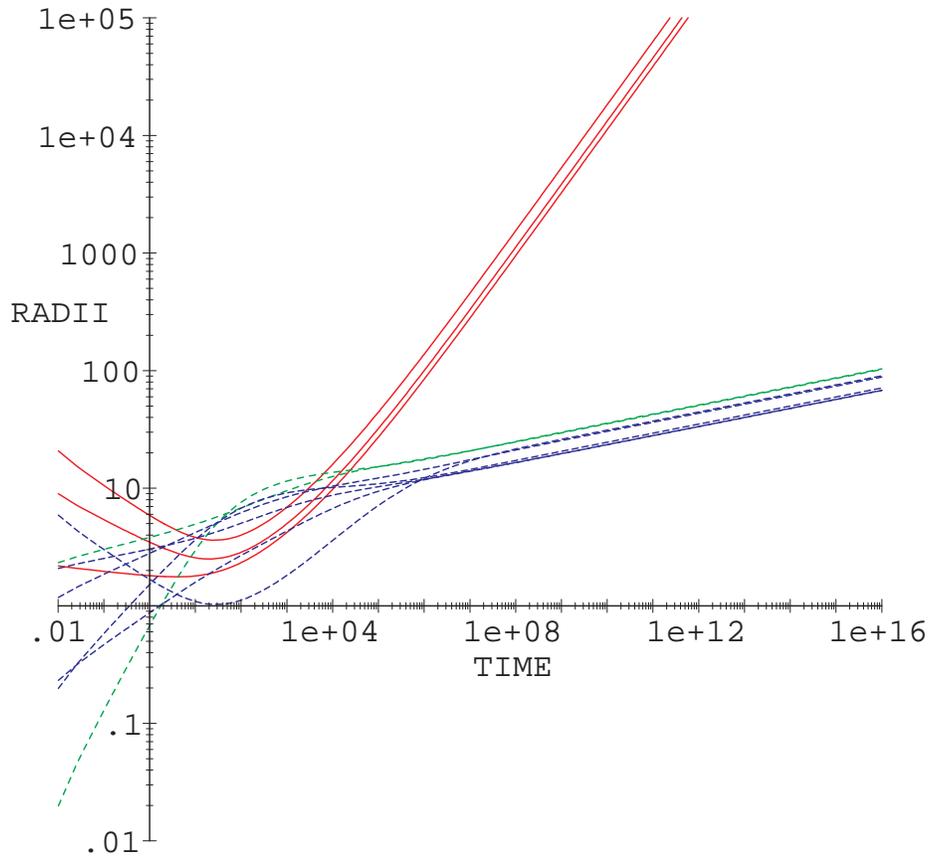}
\caption{Radii vs.~time for three unwrapped dimensions (solid red curves),
two partially wrapped dimensions (dotted green curves) and five fully wrapped
dimensions (dashed blue curves).  Distances and and times measured in Planck
units.}
\end{figure}

\begin{figure}
\epsfig{file=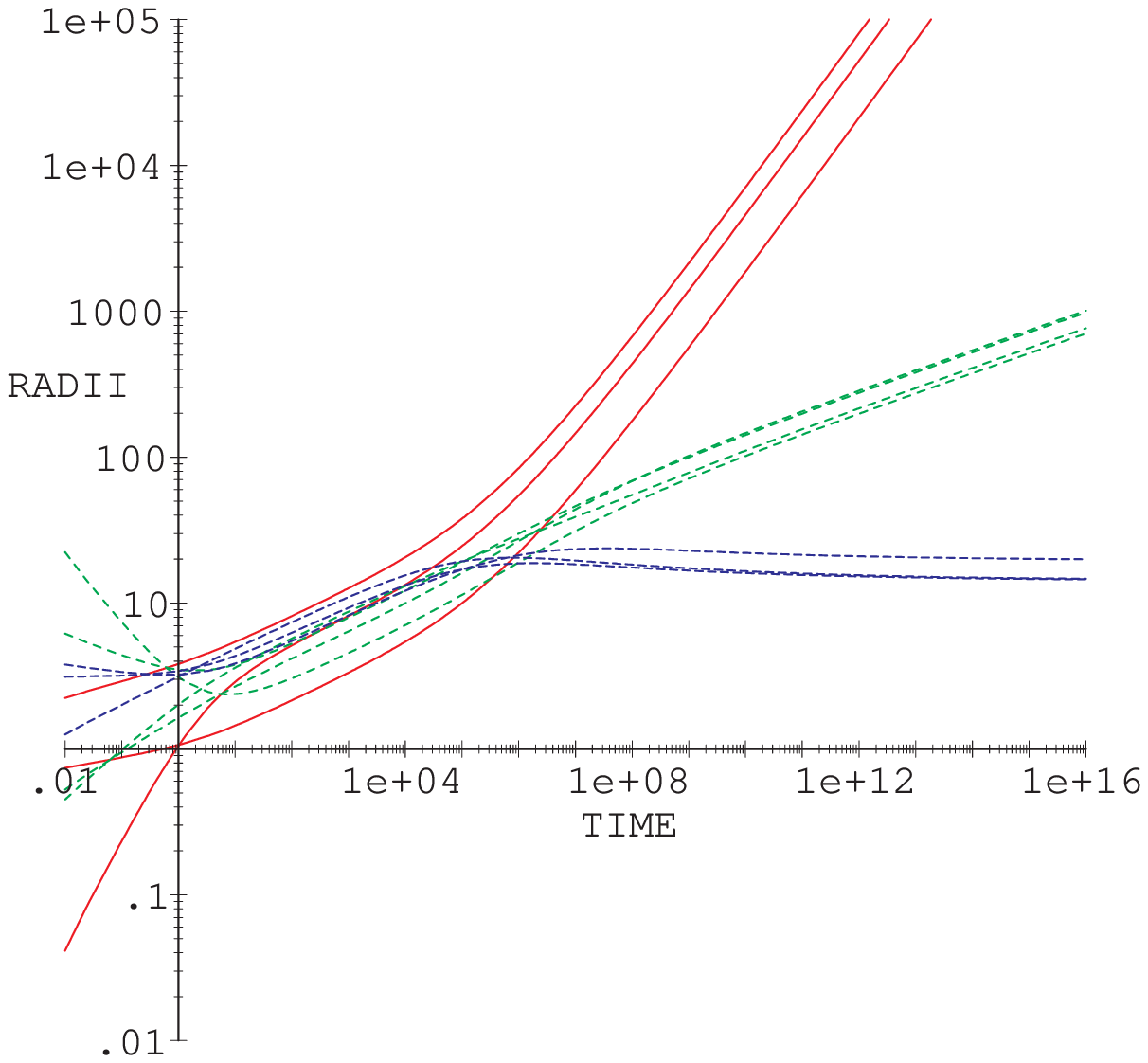}
\caption{A plot with three unwrapped dimensions (solid red curves), four
partially wrapped dimensions (dotted green curves) and three fully wrapped
dimensions (dashed blue curves).}
\end{figure}

The equations predict that the early universe is radiation-dominated,
with the volume of the spatial torus being ${\rm V} \sim
t^{2d/(d+1)}$.  Thus the universe originates in a big bang, which we
have taken to be at $t = 0$.  The big bang is apparently highly
anisotropic, with some radii increasing while others go to zero, but
we do not trust our equations of motion at early times: the
temperature diverges while many of the radii are sub-Planckian, so we
do not expect supergravity to be valid.  It would be interesting to
understand what, if anything, resolves the initial singularity in
M-theory.  Presumably U-duality plays a role, perhaps along the lines
that have been argued in string theory \cite{bv}.

We do expect supergravity to hold at late times, due to the growing
radii and falling temperature.  The growing radii allow us to neglect
brane--antibrane annihilation, which turns off as the universe
expands, and excitations on the branes will get red-shifted away
faster than the brane tension.  We therefore trust our solutions to
provide an accurate description of the system at late times,
presumably until low-energy dynamics eventually stabilize the moduli.

Thus the most significant feature of these solutions is that, after a
rather long transition period, the unwrapped radii begin to grow as a
universal power of $t$.  The partially and fully wrapped radii also
grow as universal powers of $t$, but with different (smaller)
exponents. This leads to the dynamical formation of a hierarchy
between the wrapped and unwrapped directions.  We will determine the
power laws analytically in the next section.

To conclude, let us comment on a less generic feature of these
solutions.  As we will see in the next section, the exponents
appearing in the late-time power laws depend only on the values of
$m_1$, $m_2$, $m_3$.  However the coefficients in front of the power
laws depend have complicated dependence on the wrapping numbers.  A
wrapping matrix of the form (\ref{WrappingMatrix}) tends to make the
partially wrapped dimensions bigger than the fully wrapped dimensions
at late times.  This effect can be seen in the figures, and suggests
that brane gas cosmology also leads to a sub-hierarchy in the sizes of
the internal dimensions.\footnote{Of course the unwrapped dimensions
grow with a faster power law and always become much larger than the
internal dimensions.}  In the case of most interest, this suggests
that one may end up with five small dimensions, two intermediate-sized
dimensions, and three decompactified dimensions.  Such a hierarchy has
also been argued to emerge from weakly-coupled string theory
\cite{AlexanderBrandenbergerEasson}.

The intriguing possibility of the dynamical hierarchy comes with a
number of qualifications, however.  We need a better understanding of thermal
fluctuations at early times, to determine what sort of wrapping
numbers are reasonable.  In addition, in this scenario some low-energy
mechanism must eventually stabilize the moduli, and it is quite
possible that this mechanism will erase any hierarchy that forms.

\section{Late time behavior}
\label{LateTime}

To complement our numerical results, we proceed to study the behavior
of the radii at late times analytically.  Rather remarkably, there is
an attractor mechanism at work, giving late-time behavior that is
independent of the initial radii and velocities.

We study the late-time behavior with $m_1$ unwrapped dimensions,
$m_2$ partially wrapped dimensions and $m_3$ fully wrapped dimensions.
That is, we consider wrapping matrices of the form of
(\ref{WrappingMatrix}). 

The late-time behavior of solutions to the equations of motion
(\ref{eom}) is captured by the ansatz
\be
\label{ansatz}
2 \pi R_i = e^{\lambda_i} = \left\lbrace
\begin{array}{ll}
a_i t^\alpha & i = 1,\ldots,m_1 \quad \hbox{\rm (unwrapped)} \\
b_i t^\beta &  i = m_1+1,\ldots,m_1+m_2 \quad \hbox{\rm (partially wrapped)} \\
c_i t^\gamma & i = m_1+m_2+1,\ldots,d \quad \hbox{\rm (fully wrapped)}\,.
\end{array}
\right.
\ee
With this ansatz, the energy densities due to supergravity particles and brane tension
are given by
\bea
& & \rho_{\rm S} = \left({{\rm const.} \over V}\right)^{(d+1)/d}
= \left({{\rm const.} \over \prod a_i \prod b_i \prod c_i t^{m_1 \alpha + m_2\beta + m_3
\gamma}}\right)^{(d+1)/d} \\
\nonumber
& & \rho_{\rm B} = {T_2 \over \prod a_i \prod b_i \prod c_i t^{m_1 \alpha + m_2 \beta
+ m_3 \gamma}} \times \\
 &&\mbox{}
 \qquad  \left( \sum_{i \not= j} (N_{ij}+N_{ji}) b_i c_j t^{\beta + \gamma}
+ \sum_{i \not= j} N_{ij} c_i c_j t^{2 \gamma}\right)\,.
\eea
There are several distinct classes of behavior at late times since,
depending on the values of $m_1$, $m_2$, $m_3$, certain contributions
to the energy density become negligible at late times.  The following
four subsections exhaust the possibilities.

\subsection{Neglect $\rho_{\rm S}$}
\label{Sugra}

If the universe expands rapidly enough then it is brane-dominated at
late times, with $\rho_{\rm B} \gg \rho_{\rm S}$.  We begin by
analyzing this case, which turns out to be the one of greatest
physical interest.

Let us assume that every term in $\rho_{\rm B}$ makes a comparable
contribution to the energy density.  This amounts to assuming that
$\beta = \gamma$.  Substituting the ansatz (\ref{ansatz}) into the
equations of motion (\ref{eom2}), and neglecting $\rho_{\rm S}$, the
time dependence cancels out of the equations of motion as long as
\be
\label{eqn1}
m_1 \alpha + (m_2 + m_3 - 2) \beta = 2\,.
\ee
The equation of motion for each unwrapped dimension becomes
\be
\label{eqn2}
\alpha(m_1 \alpha + (m_2 + m_3)\beta - 1) = {3 \over d-1}
8 \pi G t^2 \rho_{\rm B} \,.
\ee
On the other hand, summing the equations of motion over all partially
and fully wrapped dimensions (that is, from $i=m_1+1$ to $d$) gives
\be
\label{eqn3}
\beta (m2+m3) (m_1\alpha + (m_2+m_3)\beta - 1)
= \left({3(m_2+m_3)\over d-1} - 2\right)8 \pi G t^2 \rho_{\rm B} \,.
\ee
Comparing (\ref{eqn2}) and (\ref{eqn3}), one must have
\be
\label{eqn4}
{\alpha \over (m_2+m_3) \beta} = {3 \over d - 3m_1 + 2}\,.
\ee
The solution to (\ref{eqn1}) and (\ref{eqn4}) is
\bea
\label{BraneExponents}
\alpha & = & {6(d-m_1) \over d(d-m_1) + 4m_1 - 4} \\
\nonumber
\beta = \gamma & = & {2d - 6m_1 + 4 \over d(d-m_1) + 4m_1 - 4}\,.
\eea
Note that the exponents depend on the spatial dimension $d$ and the
number of unwrapped dimensions $m_1$, but not on $m_2$ or $m_3$
separately.

The late-time power law exponents are therefore fixed by $d$ and
$m_1$.  We now show that the coefficients in front of the power laws
are also depend on the wrapping matrix.  The ansatz (\ref{ansatz})
reduces the equations of motion to a set of algebraic equations.  The
equations of motion for the unwrapped dimensions are all identical,
and give a single nontrivial equation.  The equations of motion for
the partially and fully wrapped dimensions give another $m_2+m_3$
equations.  In addition equation (\ref{eqn1}) must be imposed.  Thus
we have a total of $m_2 + m_3 + 2$ equations.  Now let's count the
unknowns.  Since $a_i$ appears only in the combination $\prod_i a_i$,
there are a total of $m_2+m_3+3$ unknowns $\alpha$, $\beta=\gamma$,
$\prod a_i$, $b_i$, $c_i$.  Thus for generic wrapping numbers the
equations are sufficient to determine all of the unknowns in terms of
the wrapping matrix, up to a freedom to make rescalings
\be
\label{ScaleSymmetry}
a_i \rightarrow \Omega a_i \qquad b_i \rightarrow
\Omega^{-m_1/(m_2+m_3-2)} b_i
\qquad c_i \rightarrow \Omega^{-m_1/(m_2+m_3-2)} c_i
\ee
which leave the equations invariant (for example, note that $\rho_{\rm
B}$ is invariant under this rescaling).

This shows that there is an attractor mechanism at work.  Modulo the
scale symmetry (\ref{ScaleSymmetry}), the late time behavior is
determined solely by the wrapping matrix.  That is, the system forgets
about the initial values of the radii $R_i$ and velocities
$\dot{R}_i$.

Of course, this analysis assumes that we can ignore $\rho_{\rm S}$.
This is only consistent if
\be
(m_1 \alpha + m_2 \beta + m_3 \gamma)(d+1)/d > 2
\ee
or equivalently
\be
m_1 < 3d(d+2)/(7d+2)\,.
\ee
This analysis also assumes that all terms in $\rho_{\rm B}$ are
comparable.  By comparing to section \ref{SugraFull}, this requires
$m_3 > m_2$.

\subsection{Neglect $\rho_{\rm S}$ and full-full wrapping}
\label{SugraFull}

For a highly anisotropic wrapping, the partially and fully wrapped
dimensions will behave differently at late times.  To analyze this
possibility, assume that $\gamma < \beta$.  This makes the full--full
wrapping terms in the brane energy density negligible at late times.
Let's also assume that $\rho_{\rm S}$ is negligible.

We get one equation by requiring that $\rho_{\rm B} \sim 1/t^2$, namely
\be
m_1 \alpha + (m_2 - 1)\beta + (m_3 - 1)\gamma = 2\,.
\ee
The equation of motion for the unwrapped dimensions reads
\be
\alpha(m_1 \alpha + m_2 \beta + m_3 \gamma - 1) = {3 \over d-1} 8 \pi G t^2 \rho_{\rm B}
\ee
while the sum of the equations of motion for the partially wrapped dimensions gives
\be
m_2 \beta (m_1 \alpha + m_2 \beta + m_3 \gamma - 1) = \left({3 m_2 \over d-1} - 1\right)
8 \pi G t^2 \rho_{\rm B}
\ee
and for the fully wrapped
\be
m_3 \gamma (m_1 \alpha + m_2 \beta + m_3 \gamma - 1) = \left({3 m_3 \over d-1} - 1\right)
8 \pi G t^2 \rho_{\rm B} \,.
\ee
This is sufficient to fix the exponents
\beas
\alpha & = & {6 m_2 m_3 \over (d-4) m_2 m_3 + (d-1) (m_2 + m_3)} \\
\beta & = & {2 m_3(3 m_2 - d + 1) \over (d-4) m_2 m_3 + (d-1) (m_2 + m_3)} \\
\gamma & = & {2 m_2 (3 m_3 -d + 1) \over (d-4) m_2 m_3 + (d-1) (m_2 + m_3)}
\eeas
A little equation counting shows that the coefficients of the power
laws are determined by the wrapping matrix, up to the freedom to make
rescalings by
\be
a_i \rightarrow \Omega a_i \qquad b_i \rightarrow \Omega^{-m_1/(m_2+m_3-2)} b_i
\qquad c_i \rightarrow \Omega^{-m_1/(m_2+m_3-2)} c_i
\ee
and by
\be
a_i \rightarrow a_i \qquad b_i \rightarrow \Omega^{1/(m_2-1)} b_i
\qquad c_i \rightarrow \Omega^{-1/(m_3 - 1)} c_i\,.
\ee
This analysis assumes that $\gamma < \beta$, or equivalently that $m_2 > m_3$.
It also assumes that $\rho_{\rm S}$ is negligible, which is consistent if
\be
(m_1 \alpha + m_2 \beta + m_3 \gamma) (d+1)/d > 2
\ee
or equivalently
\be
{m_2 m_3 \over m_2+m_3} > {d (d-1) \over 7d+2}\,.
\ee
\subsection{Neglect full-full wrapping}
\label{Full}

Next assume that $\rho_{\rm S}$ makes an important contribution at
late times, but that $\gamma < \beta$ so that the full--full wrapping
terms in the brane energy density are negligible.

Requiring that $\rho_{\rm S}$ and $\rho_{\rm B}$ scale like $1/t^2$
gives two conditions on the exponents.  Together with the equations of
motion, this is sufficient to fix
\beas
\alpha & = & {2 \over d+1} {d(m_2+m_3) + 2 m_2 m_3 \over d(m_2+m_3) -
4 m_2m_3} \\ 
\beta & = & {2 \over d+1} {d(m_2-2m_3) + 2 m_2 m_3 \over d(m_2+m_3) -
4 m_2m_3} \\ 
\gamma & = & {2 \over d+1} {d(m_3-2m_2) + 2 m_2 m_3 \over d(m_2+m_3) -
4 m_2m_3} \\ 
\eeas
The coefficients in front of the power laws are fixed up to the
rescaling freedom
\be
a_i \rightarrow \Omega^{m_2 - m_3} a_i \qquad
b_i \rightarrow \Omega^{-m_1} b_i \qquad
c_i \rightarrow \Omega^{m_1} c_i
\ee
which leaves $\rho_{\rm S}$ and the relevant terms in $\rho_{\rm B}$ invariant.

Neglecting the full--full wrapping is consistent if $\gamma < \beta$, or equivalently
if $m_2 > m_3$.  By comparison with section \ref{SugraFull}, we need $m_2 m_3 / (m_2 + m_3)
< d(d-1)/(7d+2)$ in order for $\rho_{\rm S}$ to make an important contribution.

\subsection{All terms important} 

Finally, consider the possibility that all terms in the energy density
are important.  Plugging the ansatz (\ref{ansatz}) into the equations
of motion the time dependence cancels provided that both $\rho_{\rm
S}$ and $\rho_{\rm B}$ scale like $1/t^2$.  If we assume that all
contributions to $\rho_{\rm B}$ are equally important this fixes the
exponents
\bea
\label{OtherExponents}
\alpha & = & {3d - m_1 \over m_1(d+1)} \\
\nonumber
\beta = \gamma & = & - {1 \over d+1}\,.
\eea
Note that, just as in section \ref{Sugra}, the exponents depend on
$d$ and $m_1$, but not on $m_2$ or $m_3$ separately.

The equations of motion become a system of $m_2+m_3+1$ algebraic
equations for the $m_2+m_3+1$ unknowns $\prod a_i$, $b_i$, $c_i$.  In
this case there is a true attractor mechanism at work: the unknowns
are uniquely determined since the scale symmetry (\ref{ScaleSymmetry})
is broken.  This analysis assumes that all contributions to the energy
density are equally important at late times.  By comparing to the
other cases we analyzed, this requires $m_3 > m_2$ and $m_1 >
3d(d+2)/(7d+2)$.

\begin{table}[tb]
\centering
\begin{tabular}{|r|rrrrrrrrrrr|}
\hline
$m_1$ & 0 & 1 & 2 & 3 & 4 & 5 & 6 & 7 & 8 & 9 & 10 \\
\hline
$\alpha$ & --\rule{0pt}{15pt} & $3 \over 5$ & $4\over7$ & $7\over13$ & 
$1\over2$ & $5\over11$ & $4\over11$ & $23\over77$ &
$1\over4$ & $7\over33$ & $2\over11$ \\
$\beta,\gamma$ & $1\over4$ & $1\over5$ & $1\over7$ & $1\over13$ &
$0$ & 
$-{1\over11}$ & $-{1\over11}$ & $-{1\over11}$ & $-{1\over11}$ & 
$-{1\over11}$ &--\rule[-6pt]{0pt}{21pt}\\
\hline
\end{tabular} 
\caption{Exponents for $d=10$ spatial dimensions as a function of the
number of unwrapped dimensions $m_1$, for universes with more fully
wrapped than partially wrapped dimensions ($m_3 > m_2$).}
\end{table}

\bigskip
To summarize, the late time behavior falls into four classes,
depending on the values of $m_1$, $m_2$, $m_3$.  The various classes
are characterized by whether $\rho_{\rm S}$, certain terms in
$\rho_{\rm B}$, or both can be neglected at late times.  The exponents
for each of these cases are worked out above; the results for $d=10$
and $m_3 > m_2$ are given in Table 1.  Moreover, we have shown that
the coefficients in front of the power laws are (modulo possible scale
symmetries) uniquely determined by the wrapping matrix.  This means,
for example, that the ratios of the sizes of the fully wrapped
dimensions stay constant as the universe evolves.

One important issue remains to be addressed -- whether we expect our
equations of motion to provide an accurate description of the dynamics
at late times.  Classical supergravity is only valid if all radii are
larger than the 11-dimensional Planck length, so we should only trust
solutions in which the radii are either constant or grow with time.
For such solutions we are also justified in neglecting massless
excitations on the branes, which redshift away faster than the brane
tension.  We can also neglect brane--antibrane annihilation (which
will turn off as the transverse dimensions expand).  The condition for
growing radii is $\beta,\gamma > 0$.  This restricts us to considering
one of two cases: either the universes discussed in section
\ref{Sugra} but with
\be
\label{case1}
m_1 < {d + 2 \over 3}
\ee
or the universes discussed in section \ref{SugraFull} but with
\be
\label{case2}
m_3 > {d - 1 \over 3}\,.
\ee

\section{String gas cosmology}
\label{stringsect}

Aside from the cases (\ref{case1}) and (\ref{case2}) some of the
wrapped dimensions will shrink with time, and the supergravity
approximation to M-theory eventually breaks down.  To follow the
evolution of the universe, one must take the U-duality group of
M-theory on $T^{10}$ into account\cite{udual}.
As an example of this analysis, we study the following rather special
set-up, which will allow us to make contact with the original string
gas scenario of \cite{bv,TseytlinVafa}.  Consider a wrapping matrix in which the
only non-zero matrix elements are $N_{i,10}$ and $N_{10,i}$ for $i =
1,\ldots,9$.  Regarding $x^{10}$ as the dilaton direction, this
corresponds to type IIA string theory on $T^9$, with a gas of
fundamental strings wound around all directions of the torus.  One can
solve the equations of motion (\ref{eom}) at late times with the
following power-law ansatz for the radii.
\bea
\nonumber
& & R_i(t) \sim t^\beta \qquad i = 1,\ldots,9 \\
& & R_{10}(t) \sim t^\gamma
\eea
This analysis was performed in detail in section \ref{Full}.  One
finds that $\rho_{\rm S}$ and $\rho_{\rm B}$ make comparable
contributions to the energy density at late times.  Moreover, one
finds that $R_{10}$ shrinks with time.  This invalidates the use of an
eleven-dimensional equation of state for the supergravity gas, which
is implicitly encoded in the constraint (\ref{Constraint}), and which
would have
\be
\rho_{\rm S} \sim \left({1 \over t^{9\beta + \gamma}}\right)^{11/10}\,.
\ee
Instead the gas is effectively ten dimensional, and we should take
the energy density to scale as
\be
\rho_{\rm S} \sim {1 \over t^\gamma t^{10 \beta}}\,.
\ee
The energy from brane tension scales as
\be
\rho_{\rm B} \sim {1 \over t^{(d-2) \beta}}\,.
\ee
The equations of motion (\ref{eom2}) require that both
$\rho_{\rm S}$ and $\rho_{\rm B}$ scale like $1/t^2$, which
fixes the exponents\footnote{Although it may seem we are mixing
11-dimensional equations of motion with a 10-dimensional equation of
state, we are secretly using the fact that IIA supergravity can be
obtained by dimensional reduction from M-theory.}
\be
\beta = 1/4, \qquad \gamma = -1/2\,.
\ee

This determines the late-time behavior of the M-theory metric.  But
since $R_{10}$ is shrinking, we should really reinterpret this as a
IIA string solution, using \cite{witten}
\be
ds^2_{\rm M-theory} = e^{-2\phi/3} ds^2_{\rm string} + e^{4\phi/3}
(dx^{10})^2
\ee
where $\phi$ is the dilaton and $ds^2_{\rm string}$ is the string-frame
metric.  Thus
\bea
& & e^\phi \sim t^{-3/4}, \\
& & ds^2_{\rm string} \sim {\rm const.}
\eea
This makes contact with the scenario of \cite{bv, TseytlinVafa}, where
a gas of string winding modes leads to dimensions whose size (as
measured in the string frame) is stabilized at around the string
scale.  Note that this scenario, which proposed string winding as a
way to stabilize radii, in fact has a growing volume with respect to
the M-theory frame.

\section{Conclusions}
 
We have presented solutions to the equations of motion of M-theory in
which a hierarchy arises between the wrapped and unwrapped dimensions.
Moreover, these solutions can potentially differentiate between the
wrapped dimensions themselves, depending on the specific form of the
wrapping matrix, producing a ``sub-hierarchy'' of compact dimensions.
While these results are very suggestive, we have left a number of open
issues.

First, while the wrapped directions grow far more slowly than the
unwrapped directions, they are not actually stabilized in the most
interesting case with three unwrapped (and therefore macroscopic)
directions.  However, the fluctuations envisioned by the
Brandenberger-Vafa mechanism in a string dominated universe are
actually {\em more\/} likely to generate one or two macroscopic directions,
since these fluctuations are smaller and thus favored over those that
remove the winding strings from a three dimensional subspace.
Brandenberger and Vafa resolve this problem by pointing out that
anthropic arguments make it very unlikely that a universe with only
one or two large dimensions could actually be ``observed''.  In the
case discussed here, though, taking three completely unwrapped
dimensions results in a set of compact directions that expands
more slowly than the one or two dimensional cases. Consequently, it is
presumably easier for the three dimensional models to be stabilized
by non-perturbative effects not accounted for by the supergravity action we
are working with. Thus it is at least possible that a three
dimensional universe could eventually be picked out as a special case
where one had both a permitted number of ``large'' dimensions, and a
stabilized set of compact directions, which is vital if we are to
avoid constraints on the time dependence of Newton's
constant in the ``effective'' three dimensional universe. However, it
is not clear whether one can eliminate the anthropic argument entirely
from this scenario.

To conclude, let us mention some important directions for future work.
In this paper we have focussed on particular wrapping configurations,
which lead to cosmologically interesting evolution.  The simple
topological argument introduced by Brandenberger and Vafa suffices to
show that our wrapping configurations are {\em possible}: they could
arise from thermal fluctuations in the early universe.  However we
have not attempted to analyze the {\em likelihood} of such a
fluctuation.  We are currently studying the thermodynamics of the
early universe in this scenario, including the thermal creation and
annihilation of branes \cite{next}.  An important ingredient in this
is calculating the growth of the cross-section with wrapped radii and
temperature; this is analogous to the growth of transverse
fluctuations of a string with oscillation number and resolution time
\cite{KKS}.  This will allow us to estimate the probability of
obtaining the brane wrapping configurations we have investigated.

In the present paper we have only considered spatially homogeneous
universes.  The assumption of spatial homogeneity greatly simplifies
the analysis, but to truly understand the robustness of this scenario,
it is important to relax this assumption.  Several new effects must be
included in spatially inhomogeneous universes.  In particular the
long-range forces between between branes, mediated by the exchange of
supergravity particles, must be taken into account.

\bigskip
\goodbreak
\centerline{\bf Acknowledgements}
BG and DK are supported in part by DOE grant DE-FG02-92ER40699.  MGJ
is supported by a Pfister Fellowship and is grateful to J.~Selvaggi
for financial support.  ISCAP gratefully acknowledges the generous
support of the Ohrstrom Foundation. RE and BG thank the Aspen Center
for Physics for its hospitality, where part of this work was carried
out.  We are extremely grateful to Don Marolf for several very
valuable conversations, and for sharing his unpublished calculations
analyzing a scenario very similar to the one described in this paper.


\end{document}